\documentclass[a4paper,12pt]{article}
\usepackage{graphicx,epsf}

\begin{document}

\title{Non-Gaussianities in two-field inflation}

\author{Filippo Vernizzi$^1$ and David Wands$^2$\\
{\small {}}\\
{\small ${}^1${\it Helsinki Institute of Physics, P.O. Box 64,}}\\
{\small {\it FIN-00014 University of Helsinki, Finland}}\\
{\small {\it and}}\\
{\small ${}^2${\it Institute of Cosmology and Gravitation,
University of Portsmouth,}}\\
{\small {\it Portsmouth PO1 2EG, United Kingdom}}\\
}

\date{\today}
\maketitle

\def\beq{\begin{equation}}
\def\eeq{\end{equation}}
\newcommand{\bea}{\begin{eqnarray}}
\newcommand{\eea}{\end{eqnarray}}
\def\bi{\begin{itemize}}
\def\ei{\end{itemize}}
\def\Tdot#1{{{#1}^{\hbox{.}}}}
\def\Tddot#1{{{#1}^{\hbox{..}}}}
\def\D{{D}}
\def\d{{\delta}}
\def\T{{\bf T}}
\def\perp{n}
\def\R{{\cal K}}
\def\L{{\cal L}_u}
\def\HH{{\cal H}}
\def\W{{\cal W}}
\def\F{{\cal F}}
\def\M{{\cal M}}
\def\B{{\cal A}}
\def\Z{Z}
\def\A{A}
\def\mp{m_\phi}
\def\mc{m_\chi}
\def\mP{m_{\rm P}}
\def\bx{{\bf x}}
\def\PP{{\cal P}}
\def\bk{{\bf k}}
\def\l{\langle}
\def\r{\rangle}
\def\R{{\cal R}}
\def\N{{\cal N}}
\def\q{q}
\def\u{c}
\def\f{f}
\newcommand{\gsim}{\ \raise.3ex\hbox{$>$\kern-.75em\lower1ex\hbox{$\sim$}} \ }
\newcommand{\lsim}{\ \raise.3ex\hbox{$<$\kern-.75em\lower1ex\hbox{$\sim$}} \ }

\begin{abstract}
We study the bispectrum of the curvature perturbation on uniform
energy density hypersurfaces in models of inflation with two
scalar fields evolving simultaneously. In the case of a separable
potential, it is possible to compute the curvature perturbation up
to second order in the perturbations, generated on large scales
due to the presence of non-adiabatic perturbations, by employing
the $\delta N$-formalism, in the slow-roll approximation. In this
case, we provide an analytic formula for the nonlinear parameter
$f_{\rm NL}$. We apply this formula to double inflation with two
massive fields, showing that it does not generate significant
non-Gaussianity; the nonlinear parameter at the end of inflation
is slow-roll suppressed. Finally, we develop a numerical method
for generic two-field models of inflation, which allows us to go
beyond the slow-roll approximation and confirms our analytic
results for double inflation.
\end{abstract}

\newpage

\section{Introduction}

A key prediction of a period of inflation in the very early
universe, is the generation of a spectrum of primordial
perturbations. Such perturbations naturally arise from the zero
point vacuum fluctuations in quantum fields, which are stretched
to arbitrarily large scales during inflation. The distribution of
the primordial density perturbations thus provides an important
test of any inflation model. In particular, slow-roll models of
inflation generically predict an almost scale-invariant, almost
Gaussian distribution of primordial density perturbations
\cite{Lyth:1998xn,liddle_lyth}.

Typically one calculates the power spectrum of density
perturbations and the relative amplitude of gravitational waves.
However, there is limited information in the power spectrum over
the restricted range of scales where the primordial power spectrum
can be reliably inferred from observations. As a result, there has
been growing interest in calculating the distribution of
primordial perturbations from inflation, considering not only the
power spectrum but also the bispectrum and other measures of
possible deviations from purely Gaussian distribution as a
possible discriminant between different models
\cite{Bartolo:2004if,Babich:2004gb}.  For instance, it is known
that the size of the bispectrum during single-field, slow-roll
inflation is related to the spectral tilt of the power spectrum
and is thus constrained to be small \cite{Maldacena}. On the other
hand, inflationary models with higher derivative operators, such
as ghost inflation or inflation based on the Dirac-Born-Infeld
action, can produce higher non-Gaussianity
\cite{Arkani-Hamed:2003uz,Alishahiha:2004eh,Seery,Creminelli},
possibly detectable. Furthermore, non-adiabatic perturbations
produced during {\em multi-field} inflation can certainly generate
detectable non-Gaussianity in the density field {\it after}
inflation, in models such as the curvaton scenario
\cite{Lyth:2002my} (see also \cite{JP}) or reheating
\cite{Zaldarriaga:2003my,Enqvist:2004ey,Barnaby:2006cq,Jokinen:2005by}.

What is less clear is whether nonlinear large-scale evolution of
perturbations {\em during} inflation is capable of producing
significant non-Gaussianity. Such a study requires a consistent
treatment of nonlinear perturbations, not just in the matter
fields but also in the gravitational field. Recently a number of
authors have developed gauge-invariant descriptions of the
nonlinear curvature perturbation on large scales
\cite{Malik:2003mv,Rigopoulos:2003ak,Lyth:2004gb,Vernizzi:2004nc,Lyth:2005fi,Langlois:2005ii}.

Rigopoulos {\em et al.}~\cite{Rigopoulos:2005us} have used
numerical simulations of the stochastic dynamics on large scales
and found significant non-Gaussianity even in simple two-field
models. It is this question which we wish to address in this paper
by presenting analytical and numerical estimates of the
non-Gaussianity in two-field inflation models. We use the $\delta
N$-formalism, as advocated by Lyth and Rodriguez
\cite{Lyth:2005fi}, to calculate the evolution of the curvature
perturbation after Hubble exit. We find that in the case of a
separable potential, it is possible to construct an analytical
formula to express the bispectrum of the curvature perturbation in
terms of the potentials and slow-roll parameters of the two
fields. Furthermore, we also use numerical solutions to go beyond
the slow-roll approximation after Hubble exit, confirming our
analytic results. In the case of double inflation with two massive
fields, we find no significant non-Gaussianity produced, which
confirms a previous discussion by Alabidi and Lyth in
Ref.~\cite{Alabidi}.

Note that the multi-field scenario considered here is different
from that considered in \cite{JP,Enqvist:2004bk}, where only one
field is contributing to the energy density during inflation.

In Sec.~2 we review the two- and three-point statistics of field
perturbations during slow-roll inflation, and how in the $\delta
N$-formalism these can be related to the two- and three-point
statistics of the curvature perturbation on uniform density
hypersurfaces in arbitrary multi-field models. In Sec.~3 we show
how one can actually calculate the evolution of the primordial
power spectrum and bispectrum using the slow-roll approximation in
simple two-field models with a separable potential, during
inflation. In Sec.~4 we discuss the issue of how the end of
inflation can affect the curvature perturbation and the
non-Gaussianity. In Sec.~5 we present both an analytical and
numerical study of non-Gaussianity in the double quadratic
inflation model studied by Rigopoulos {\em et
al.}~\cite{Rigopoulos:2005us}, going beyond the slow-roll
approximation. We present our conclusions in Sec.~6.

\section{Non-Gaus\-sia\-n perturbations in multi-field inflation}

Here we review some of the results concerning perturbations and
non-Gaus\-sia\-ni\-ties in multi-field inflationary models. For
simplicity, we will consider a model of inflation driven by a set
of minimally coupled scalar fields with canonical kinetic terms
and potential $W$, described by the action
\beq
S= - \frac{\mP^2}{2 } \int d^4 x \sqrt{-g} \left[ \frac{1}{2}
\sum_I \partial^\mu \varphi_I
\partial_\mu \varphi_I + W(\varphi_1, \varphi_2, \ldots)
\right],
\eeq
where $\mP^2\equiv8 \pi G$ is the reduced squared Planck mass.

We consider scalar perturbations in a quasi-homogeneous flat
Friedmann-Lema\^{\i}tre-Robertson-Walker spacetime with
scale factor $a(t)$ and perturbed metric
\beq
ds^2 = -(1+2 A) dt^2 + 2 B_{,i} dt dx^i + a(t)^2
\left[(1-2\psi)\delta_{ij} + 2 E_{,ij} \right] dx^i dx^j.
\eeq

Primordial cosmological perturbations are usually expressed in
terms of the curvature perturbation on uniform energy density
hypersurfaces, denoted by $\zeta$, defined in
\cite{Bardeen:1980kt,Bardeen:1983qw} in linear perturbation
theory, and generalized at higher order in the perturbations in
\cite{Malik:2003mv,Rigopoulos:2003ak,Lyth:2004gb}. (See also
\cite{Vernizzi:2004nc,Lyth:2005du,Malik:2005cy} for comparison
between different variables at second order.) This quantity is
widely used, especially because it is conserved, on large scales,
for adiabatic perturbations \cite{Wands:2000dp}. Here we want to
compute the three-point correlation function of $\zeta$ and hence
we need to define $\zeta$ only up to {\em second order} in the
perturbations,
\beq
\zeta \equiv -\psi -\psi^2 - {H\over {\dot \rho}} \delta \rho
 + \frac{1}{{\dot \rho}} \dot
\psi \delta \rho +  \frac{H}{\dot \rho^2} \delta \rho
 \dot{\delta \rho} +\frac{1}{2 \dot \rho}
 \Tdot{\left({H \over \dot \rho }\right)}
 {\delta \rho}^2,
\eeq
where $\rho$ is the background energy density, $\delta \rho$ its
perturbation, and $H\equiv \dot a /a$ is the Hubble rate, while a
dot denotes the derivative with respect to cosmic time $t$.

According to the so called $\delta N$-formalism
\cite{Starobinski,Sasaki:1995aw,Sasaki:1998ug}, $\zeta$, evaluated
at some time $t=t_\u$, is equivalent, {\em on large scales}, to
the perturbation of the number of $e$-foldings $\N(t_\u, t_*,
\bx)$ from an initial {\em flat} hypersurface at $t=t_*$, to a
final {\em comoving} -- or, equivalently, {\em uniform density} --
hypersurface at $t=t_\u$. Hence, one has, on large scales,
\beq
\zeta(t_\u,\bx) \simeq \delta N(t_\u, t_*, \bx) \equiv \N(t_\u,
t_*, \bx) - N(t_\u, t_*), \label{zeta}
\eeq
where
\beq
N(t_\u, t_*) \equiv \int_*^\u H dt \label{N}
\eeq
is the unperturbed value of $\cal N$. Equation (\ref{zeta}) simply
follows from the definition of $\N$ as the volume expansion rate
of the $t=$ const hypersurface, integrated along the integral
curve of the unit vector orthogonal to the $t=$ const hypersurface
\cite{Sasaki:1995aw}. This definition is not restricted to linear
theory but holds also at second or higher order in the
cosmological perturbations
\cite{Sasaki:1998ug,Lyth:2003im,Lyth:2004gb,Langlois:2005ii}.

One can then take $t_*$ as the time, during inflation, when the
relevant perturbation scales exited the Hubble radius, $k=aH$, and
$t_\u$ as some time $> t_*$ during or after inflation. Then the
number of $e$-foldings $\N$ can be viewed as a function of the
field configuration $\varphi_I(t_*,\bx)$ on the flat hypersurface
at $t=t_*$ and of the time $t_\u$. If one splits the scalar fields
in a background value and a perturbation, $\varphi_I(t, \bx)
\equiv \varphi_I(t)+ \delta \varphi_I(t,\bx)$, $\delta N(t_\u,
t_*, \bx)$ can be expanded in series of the initial field perturbations
$\delta \varphi_I(t_*,\bx)$. Retaining only terms up to second
order, one obtains
\beq
\delta N(t_\u,t_*,\bx)= \sum_I N_{,I} \delta \varphi_*^I + \frac{1}{2}
\sum_{IJ} N_{,IJ} \delta \varphi_*^I \delta \varphi_*^J, \label{deltaN2}
\eeq
where
\beq
N_{,I} \equiv \frac{\partial N }{ \partial \varphi_*^I}, \qquad
N_{,IJ} \equiv \frac{\partial^2 N }{ \partial \varphi_*^I \partial
\varphi_*^J},
\eeq
are the first and second derivatives of the {\em unperturbed}
number of $e$-foldings $N(t_\u,t_*)$, with respect to the
unperturbed values of the fields at Hubble crossing. Note that in
general $\N$ depends on the fields, $\varphi_I(t)$, and their first time
derivatives, $\dot \varphi_I(t)$. However, if
slow-roll conditions
\beq
3H \dot \varphi_I \simeq - W_{,I}, \quad {\rm at } \ \  t=t_*
\label{srdN}
\eeq
are satisfied at Hubble exit,
then  $\N$ depends only on the field values \cite{Sasaki:1995aw}.

In \cite{Lyth:2005fi} it was shown that Eq.~(\ref{zeta}) with
(\ref{deltaN2}) can be used to compute $\zeta$ up to second order
in the perturbations in multi-field models of inflation. In
particular, the $\delta N$-formalism is {\em equivalent} to
integrate the evolution of $\zeta$ on super-Hubble scales from
Hubble exit until $t_\u$.

\subsection{Two-point statistics}

The power spectrum of the curvature perturbation
$\zeta$, $\PP_\zeta$, is defined as
\beq
\langle \zeta_{\bk_1} \zeta_{\bk_2} \rangle \equiv (2 \pi)^3
\delta^{(3)} (\bk_1 + \bk_2) \frac{2\pi^2}{k_1^3} \PP_\zeta (k_1) .
\label{dps}\label{ps}
\eeq

Let us now take the scalar field perturbations $\delta \varphi_I$
as uncorrelated stochastic variables at early times, with scale
invariant spectrum of massless scalar fields in de-Sitter space.
Their two-point correlation functions hence satisfy
\beq \langle
\delta\varphi^I_{\bk_1} \delta\varphi^J_{\bk_2} \rangle = (2
\pi)^3 \delta_{IJ} \delta^{(3)} (\bk_1 +
\bk_2)  \frac{2 \pi^2}{k_1^3} \PP_* (k_1) , \qquad \PP_*(k)  \equiv \frac{H_*^2}{4
\pi^2}, \label{sfps}
\eeq
where
$H_*$ is evaluated at Hubble exit, $k=aH$.
From Eqs.~(\ref{ps}) and (\ref{sfps}),
making use of Eqs.~(\ref{zeta}) and (\ref{deltaN2}),
one obtains
\beq
\PP_{\zeta}=\sum_I N_{,I}^2 \PP_*. \label{Rtophi}
\eeq

Another interesting observable that can be derived from the
two-point statistics is the scalar spectral index of $\PP_{\zeta}$,
defined as
\beq
n_\zeta-1 \equiv \frac{d \ln \PP_{\zeta}}{d \ln k}.
\eeq
For multi-field inflation, the expression for the scalar spectral
index has been given, for instance, in
\cite{Sasaki:1995aw,Lyth:1998xn}. At lowest order in slow-roll one
has \cite{liddle_lyth}
\beq
\frac{d \ln \PP_\zeta}{d \ln k} \simeq  \frac{d \ln \PP_\zeta}{d N} =
\frac{1}{H}\frac{d \ln \PP_\zeta}{d t}.
\eeq
The first equality follows from taking $k$ at Hubble crossing,
$k=aH \propto H \exp N $, while the second is a consequence of the
definition of $N$, Eq.~(\ref{N}). Making use of the expression for
the scalar power spectrum $\PP_\zeta$ given in (\ref{Rtophi}),
and of
\beq
\frac{d}{dt}= \sum_I \dot \varphi_I \frac{\partial}{\partial
\varphi_I},
\eeq
one thus obtains
\beq
n_\zeta-1= - 2 \epsilon+ \frac{2}{H} \frac{\sum_{IJ} \dot \varphi_J
N_{,JI} N_{,I}}{\sum_K N_{,K}^2}, \label{si}
\eeq
where $\epsilon$ is the well known slow-roll parameter defined as
\beq
 \epsilon \equiv - \frac{\dot H}{H^2} \,. \label{eps}
\eeq
This expression can be shown to be equivalent to that given in
\cite{Sasaki:1995aw}. Indeed, the particular combination of
second derivatives, $\dot\varphi_JN_{,IJ}$, that appears in the
spectral index (\ref{si}) can be eliminated in favor of second
derivatives of the potential using the slow-roll approximation to
give \cite{Lyth:1998xn}
\beq
n_\zeta-1 = -2\epsilon - \frac{2}{\mP^2\sum_K N_{,K}^2}
 + \frac{2\mP^2\sum_{IJ}V_{,JI}N_{,I}N_{,J}}{V\sum_K N_{,K}^2} \,.
\eeq

\subsection{Three-point statistics}

Let us now discuss the three-point statistics of the curvature
perturbations. The bispectrum of the curvature perturbation
$\zeta$ is defined as
\beq
\l \zeta_{\bk_1} \zeta_{\bk_2} \zeta_{\bk_3} \r\equiv (2 \pi)^3
\delta^{(3)}(\sum_i \bk_i) B_\zeta (k_1,k_2,k_3).
\label{bispectrum}
\eeq
Observational limits are usually put on the so called {\em
nonlinear parameter} $f_{\rm NL}$. We use here the definition of
$k$-dependent $f_{\rm NL}$ given in
\cite{Maldacena},\footnote{Note that this choice of
  sign for $f_{\rm NL}$
  differs from that used by
  \cite{KS,WMAP1}.
 The minimal detectable $f_{\rm NL}$
  with an ideal CMB experiment is slightly larger than unity \cite{Verde,KS}.}
\beq
-\frac{6}{5}f_{\rm NL} \equiv \frac{\Pi_i k_i^3}{\sum_i k_i^3}
\frac{B_\zeta}{4 \pi^4 \PP_\zeta^2}. \label{def_fNL}
\eeq

One can compute the bispectrum
using Eqs.~(\ref{zeta}) and (\ref{deltaN2}).
For the three-point correlation function of $\zeta$, this
yields
\bea
\langle \zeta_{\bk_1} \zeta_{\bk_2} \zeta_{\bk_3} \rangle &=&
\sum_{IJK} N_{,I} N_{,J} N_{,K} \langle \delta \varphi^I_{\bk_1}
\delta \varphi^J_{\bk_2} \delta \varphi^K_{\bk_3}\rangle + \nonumber
\\ &&  \frac{1}{2} \sum_{IJKL} N_{,I} N_{,J} N_{,KL} \langle \delta
\varphi^I_{\bk_1} \delta \varphi^J_{\bk_2} (\delta \varphi^K \star
\delta \varphi^L)_{\bk_3}\rangle
+{\rm perms}, \nonumber \\ &&
\label{tpf}
\eea
where a star denotes the convolution and we have neglected
correlation functions higher than the four-point (see
\cite{Seery:2005gb,Zaballa:2006pv}).

The first line represents the contribution from the three-point
correlation functions of the fields. For purely Gaussian fields,
this vanishes. However, Seery and Lidsey \cite{Seery:2005gb} have
found that during slow-roll inflation the field three-point
correlation functions do not vanish and read
\beq
\langle \delta \varphi^I_{\bk_1} \delta \varphi^J_{\bk_2} \delta
\varphi^K_{\bk_3}\rangle = (2\pi)^3 \delta^{(3)}(\sum_i \bk_i)
\frac{4 \pi^4}{\Pi_i k_i^3} \PP_*^2 \sum_{\rm perms} \frac{\dot
\varphi_I \delta_{JK}}{4 H \mP^2}  \M, \label{ftpf}
\eeq
with
\beq
\M = \M(k_1,k_2,k_3) \equiv - k_1 k_2^2 -4 \frac{k_2^2 k_3^2}{k_t} +
\frac{1}{2} k_1^3 + \frac{k_2^2 k_3^2}{k_t^2}(k_2-k_3)~,
\label{littlef}
\eeq
where $k_t=k_1+k_2+k_3$, and it is assumed that all the $k_i$ are of
the same order of magnitude so that they cross the Hubble radius
approximately at the same time. The function $\M$,
which has been written in a slightly different form from that in
Ref.~\cite{Seery:2005gb},
parameterizes the $k$-dependence of the three-point functions. The
sum is over all simultaneous rearrangements of the indices $I$,
$J$, and $K$, and the momenta $k_1$, $k_2$, and $k_3$ in $\M $,
such that the relative position of the $k_i$ is respected
\cite{Seery:2005gb}.

The sum
over the permutations of $\M $ over all the $k_i$
reads
\bea
\F(k_1,k_2,k_3) &\equiv& \sum_{{\rm perms}}
\M(k_1,k_2,k_3) \nonumber \\
&=& -2 \left(\frac{1}{2} \sum_{i\neq j} k_i
k_j^2 + 4 \frac{\sum_{i>j} k_i^2 k_j^2 }{k_t} - \frac{1}{2} \sum_i
k_i^3 \right), \label{bigf}
\eea
and one can sum over the permutations and evaluate the first line
of Eq.~(\ref{tpf}) using Eq.~(\ref{ftpf}) and
\beq
\sum_{IJK} N_{,I} N_{,J} N_{,K}  \sum_{\rm perms}   \dot \varphi_I
\delta_{JK} \M (k_1,k_2,k_3)  = - H \sum_{I} N_{,I}^2
\F(k_1,k_2,k_3),\label{sum}
\eeq
where we have used $\sum_I N_{,I} \dot \varphi_I=-H$.

To evaluate the second line of Eq.~(\ref{tpf}), as in  \cite{Seery:2005gb}
we assume that the connected part of the four-point correlation functions
is negligible
and we make use of Wick's theorem to reduce the four-point functions
to products of
two-point functions. Hence, we can finally rewrite the bispectrum
using its definition (\ref{bispectrum}) and  Eq.~(\ref{tpf}).
This yields, after few manipulations,
\beq
B(k_1,k_2,k_3) = 4 \pi^4 \PP_\zeta^2  \frac{\sum_i k_i^3}{\Pi_i
k_i^3} \left(  \frac{- 1}{4 \mP^2
\sum_{K} N_{,K}^2} \frac{\F}{\sum_i k_i^3} + \frac{\sum_{IJ}
N_{,I} N_{,J} N_{,IJ}}{(\sum_{K} N_{,K}^2)^2}\right).
\label{Robs2}
\eeq
To derive this expression we have used Eq.~(\ref{Rtophi}) to
replace $\PP_*^2$ with $\PP_\zeta^2$.

One can relate the bispectrum to the nonlinear parameter $f_{\rm
NL}$ by using Eq.~(\ref{def_fNL}),
\beq
-\frac{6}{5}f_{\rm NL} =
\frac{\PP_*}{2  \mP^2\PP_\zeta}
(1+f)
 + \frac{\sum_{IJ}
N_{,I} N_{,J} N_{,IJ}}{(\sum_{K} N_{,K}^2)^2} , \label{fNLfinal}
\eeq
where
\beq
f=f(k_1,k_2,k_3) \equiv - 1 - \frac{\F}{2
\sum_i k_i^3} \label{f_def}
\eeq
is a function of the shape of the momentum triangle with the range
of values $0\le f \le \frac{5}{6}$ \cite{Maldacena}, and we have
used Eq.~(\ref{Rtophi}) in the first term on the right hand side
of Eq.~(\ref{fNLfinal}). The lower bound on $f$ is obtained in the
geometrical limit when one of the three $k_i$'s is much smaller
than the other two, e.g., $k_1 \ll k_2 \approx k_3$, in which case
$f_{\rm NL}$ becomes independent of $k$
\cite{Maldacena,Allen:2005ye}, while the upper bound is obtained
in the equilateral triangle configuration, $k_1 \approx k_2
\approx k_3$.

The first term on the right hand side of Eq.~(\ref{fNLfinal}) is
momentum dependent and comes from the three-point correlation
functions of the fields, Eq.~(\ref{ftpf}), computed by quantizing
the perturbations {\em inside} the Hubble radius during inflation
\cite{Seery:2005gb,Zaballa}. Notice that $\PP_*$ can be related to
the amplitude of gravitational waves. Introducing the ratio
between tensor to scalar modes,
\beq
r\equiv \frac{8 \PP_*}{\mP^2 \PP_\zeta}, \label{r}
\eeq
this term can be written as
\beq
-\frac{6}{5}f^{(3)}_{\rm NL} \equiv  \frac{r}{16}
(1+f), \label{fNL3}
\eeq
and is constrained by observations to be small, $r/16\ll 1$.
This expression simplifies the bound on this term found in
Ref.~\cite{Zaballa}.

The second term \cite{Lyth:2005fi},
\beq
 \label{fNL4}
-\frac{6}{5}f^{(4)}_{\rm NL} \equiv  \frac{\sum_{IJ} N_{,I} N_{,J}
N_{,IJ}}{(\sum_{K} N_{,K}^2)^2},
\eeq
is momentum independent and {\em local} in real space, because it is due to
the evolution of nonlinearities {\em outside}
the Hubble radius during inflation.

The nonlinear parameter $f_{\rm NL}$ can only be larger than unity
if the momentum independent term $f^{(4)}_{\rm NL}$ is large. We
will devote the next section to compute this quantity in two-field
models.

\section{Two-field inflation with separable potential}

We will consider now two scalar fields $\varphi_1 \equiv \phi$ and
$\varphi_2 \equiv \chi$,
described by the action
\beq
 S= - \frac{\mP^2}{2} \int d^4 x \sqrt{-g} \left[ \frac{1}{2}
\partial^\mu \phi
\partial_\mu \phi +\frac{1}{2}
\partial^\mu \chi
\partial_\mu \chi + W(\phi, \chi)
\right] \,.
 \eeq
We assume that the potential of the fields is separable into the
sum of two potentials each of which is dependent on only one of
the two fields,
\beq
 W(\phi, \chi) = U(\phi) + V(\chi) \,. \label{potential}
\eeq

\subsection{Slow-roll dynamics of background fields}

The Klein-Gordon equations for the background fields
read
\bea
\ddot \phi+3H\dot \phi +U'=0,\label{evol_phi}\\
\ddot \chi+3H\dot \chi +V'=0, \label{evol_chi}
\eea
where
\beq
U' \equiv \frac{dU}{d \phi} ,
\qquad V' \equiv \frac{dV}{d \chi} .
\eeq
The unperturbed Friedmann equations read
\bea
H^2 &=& \frac{1}{3 \mP^2} \left( \frac{1}{2} \dot \phi^2 +
\frac{1}{2} \dot \chi^2 +W  \right), \label{fried_1}
\\
\dot H &=&-\frac{1}{2 \mP^2} \left(\dot \phi^2 + \dot \chi^2
\right). \label{fried_2}
\eea

Inflation takes place when $\epsilon = - {\dot H}/{H^2} < 1$. Here
it is assumed that the slow-roll conditions are satisfied for both
fields during inflation. In this case, the Klein-Gordon and the
first Friedmann equations reduce to
\beq
 3 H \dot \phi \simeq- U', \quad \quad 3 H
\dot \chi \simeq - V' , \quad \quad 3 \mP^2 H^2 \simeq W.
\label{slow_roll}
 \eeq
It is hence convenient to extend the
definition of slow-roll parameter $\epsilon$ for one field  and
define \cite{Garcia-Bellido:1995kc}
\beq
  \epsilon^\phi \equiv \frac{\mP^2}{2} \left(\frac{U'}{W} \right)^2, \quad \quad
  \epsilon^\chi \equiv \frac{\mP^2}{2} \left(\frac{V'}{W} \right)^2,
 \label{epsfields}
\eeq
with
\beq
 \epsilon = \epsilon^\phi + \epsilon^\chi.
 \eeq
One can also define the two slow-roll
parameters
\beq
  \eta^\phi \equiv \mP^2 \frac{U''}{W} , \quad \quad
  \eta^\chi \equiv \mP^2 \frac{V''}{W}. \label{etafields}
 \eeq
Note that throughout this paper, for simplicity, we will assume
$U'\geq0$ and $V'\geq0$ so that we may eliminate first derivatives
of the potential in favor of the slow-roll parameters
$\sqrt{\epsilon^\phi}$ and $\sqrt{\epsilon^\chi}$.

We will use the slow-roll equations (\ref{slow_roll}) to write the
number of $e$-foldings (\ref{N}) during inflation as
\cite{Starobinski,Lyth:1998xn}
\beq
 \label{Ndefinite}  \label{Nphichi}
N= - \frac{1}{\mP^2} \int^\u_*  \frac{U}{U'} d\phi -
\frac{1}{\mP^2} \int^\u_* \frac{V}{V'} d\chi.
 \eeq

There is a crucial difference between single field inflation and
inflation with many fields \cite{Garcia-Bellido:1995kc}. In
single field inflation the slow-roll solution forms a
one-dimensional phase space. This means that once the inflationary
attractor has been reached, there is a {\em unique} trajectory. In
particular, the end of inflation takes place at a fixed value of
the inflaton field which in turn corresponds to a fixed energy
density. However, if two fields are present, the phase-space
becomes two-dimensional and there is an infinite number of
possible classical trajectories in field space. The values of the
two fields at the end of inflation will in general depend on the
choice of trajectory. Interestingly,
for the potential (\ref{potential}), under
the slow-roll conditions (\ref{slow_roll}), there exists a
dimensionless integral of motion $C$, using which one can label
each slow-roll classical trajectory,
\beq
 C\equiv -\mP^2 \int \frac{d \phi}{U'} + \mP^2 \int
  \frac{d \chi}{V'}  .
 \label{integral}
 \eeq
The number of $e$-foldings $N$ in Eq.~(\ref{Nphichi})
characterizes the evolution {\em along} a given trajectory, while
the quantity $C$ allows us to parameterize motion {\em off} the
classical trajectory, and it will turn out to be very useful in
computing the curvature perturbation during inflation. In this
respect, the study presented here is very similar to the one
discussed in Ref.~\cite{Garcia-Bellido:1995kc}, in the case of a
potential that is a separable product rather than a separable sum.

\subsection{Perturbed expansion during slow-roll}
\label{sec:pesr}

In order to calculate $\zeta$, given in Eq.~(\ref{zeta}), we need
to calculate the perturbed expansion due to the field quantum
fluctuations on an initial spatially {\em flat} slice up to a
final {\em comoving} or {\em uniform density} hypersurface at
$t_\u$. To calculate the power spectrum of $\zeta$ (\ref{Rtophi})
we require only the first derivative of the expansion with respect
to the initial field values. But to calculate the scalar spectral
index (\ref{si}) or the three-point correlation function
(\ref{tpf}) we need to compute the perturbation in the number of
$e$-foldings expanded up to {\em second order} in the initial
field perturbations. We will thus proceed as follows. We will
first compute the first derivative of $N(t_\u, t_*)$ with respect
to the fields, i.e., $N_{,\phi}$ and $N_{, \chi}$, by
differentiating $N(t_\u, t_*)$ in Eq.~(\ref{Nphichi}) in $d
\phi_*$ and $d \chi_*$. With these two first derivatives we will
be able to compute the second derivatives $N_{,IJ}$.

We first note that each of the integrals in Eq.~(\ref{Ndefinite})
depends upon both $\phi_*$ and $\chi_*$. For instance, the value
of the integral over $\phi$ depends upon $\chi_*$ through the
dependence of the limit $\phi_\u$ upon $C(\phi_*, \chi_*)$, the
integral of motion (\ref{integral}) labelling the classical
trajectory. In other words one has
\bea
dN= \frac{1}{\mP^2} \left[ \left(\frac{U}{U'} \right)_* -
\frac{\partial \chi_\u}{\partial \phi_*} \left(\frac{V}{V'}
\right)_\u - \frac{\partial \phi_\u}{\partial \phi_*}
\left(\frac{U}{U'} \right)_\u
\right] d\phi_* \nonumber \\
+ \frac{1}{\mP^2} \left[ \left(\frac{V}{V'} \right)_* -
\frac{\partial \chi_\u}{\partial \chi_*} \left(\frac{V}{V'}
\right)_\u - \frac{\partial \phi_\u}{\partial \chi_*}
\left(\frac{U}{U'} \right)_\u \right] d\chi_*. \label{deltaN}
 \eea

In order to compute the derivatives of the number of $e$-foldings
with respect the initial fields $\phi_*$ and $\chi_*$, one needs
to compute $\partial \chi_\u /
\partial \chi_*$, $\partial \chi_\u / \partial \phi_*$, etc. Since
the value of $\phi_\u$ and $\chi_\u$ on a given classical
trajectory is a function of the conserved quantity $C$, one has
\bea
d\phi_\u = \frac{d \phi_\u}{dC} \left( \frac{\partial C}{\partial
\phi_*} d\phi_*+ \frac{\partial C}{\partial \chi_*} d\chi_*
\right), \nonumber \label{dependence1}\\
d\chi_\u = \frac{d \chi_\u}{dC} \left( \frac{\partial C}{\partial
\phi_*} d\phi_*+ \frac{\partial C}{\partial \chi_*} d\chi_*
\right). \label{dependence2}
\eea
Making use of Eq.~(\ref{integral}), one can easily compute
\beq
\frac{\partial C}{\partial \phi_*} = -\frac{\mP^2}{ U_*'}, \quad
\quad \frac{\partial C}{
\partial \chi_*} = \frac{\mP^2}{ V_*'} . \label{C_phi_chi}
 \eeq

We now require that $t=t_\u$ coincides with a $\rho =$ const
hypersurface, which during slow-roll is given by
\beq
 U(\phi_\u) + V(\chi_\u)={\rm const} \,.
 \eeq
Differentiating this condition yields
\beq
\frac{d \phi_\u}{dC} U_\u' +\frac{d \chi_\u}{dC} V_\u' =0.
\label{const}
\eeq
Furthermore, differentiating Eq.~(\ref{integral}) with respect to
the trajectory $C$ and using Eq.~(\ref{const}), one finally obtains
\bea
 \label{d1dC}
\mP^2 \frac{d \phi_\u}{dC} &=& - \left[ U_\u'
\left(\frac{1}{V_\u'{}^2 }
  + \frac{1}{U_\u'{}^2}
\right) \right]^{-1}, \nonumber \\
\mP^2 \frac{d \chi_\u}{dC} &=& \left[ V_\u'
\left(\frac{1}{V_\u'{}^2}
  + \frac{1}{U_\u'{}^2}
\right) \right]^{-1}.
\eea

Thus, substituting Eqs.~(\ref{C_phi_chi}) and (\ref{d1dC})
in Eq.~(\ref{dependence2}), one finds
\bea
\frac{\partial \phi_\u}{\partial \phi_*} = \frac{W_\u}{W_*}
\frac{\epsilon_\u^\chi}{\epsilon_\u} \left(
\frac{{\epsilon_\u^\phi }}{\epsilon_*^\phi} \right)^{1/2}
  , \quad \quad \frac{\partial
\phi_\u}{\partial \chi_*} = - \frac{W_\u}{W_*}
\frac{\epsilon_\u^\chi }{\epsilon_\u } \left(
\frac{{\epsilon_\u^\phi }}{\epsilon_*^\chi} \right)^{1/2}
, \nonumber \\
\frac{\partial \chi_\u}{\partial \phi_*} = - \frac{W_\u}{W_*}
\frac{\epsilon_\u^\phi}{\epsilon_\u} \left( \frac{{
\epsilon_\u^\chi }}{\epsilon_*^\phi} \right)^{1/2}
 , \quad \quad \frac{\partial
\chi_\u}{\partial \chi_*} = \frac{W_\u}{W_*}
\frac{\epsilon_\u^\phi}{\epsilon_\u} \left(
\frac{{\epsilon_\u^\chi }}{\epsilon_*^\chi} \right)^{1/2} ,
\label{dphiIofdphiJ}
\eea
where $\epsilon$ is the overall slow-roll parameter given in
Eq.~(\ref{eps}) and we have written the derivatives of the
potentials that appear in Eqs.~(\ref{C_phi_chi}) and (\ref{d1dC})
in terms of the slow-roll parameters defined in
Eq.~(\ref{epsfields}).

One can now evaluate the derivatives of the number of $e$-foldings with
respect to the initial fields $\phi_*$ and $\chi_*$, using
in Eq.~(\ref{deltaN})
the expressions (\ref{dphiIofdphiJ}).
For the first derivative this yields
\bea
\mP \frac{\partial N}{\partial \phi_*}&=&
  \frac{1}{\sqrt{2\epsilon^\phi_*}} \frac{U_*+\Z_\u}{W_*}  , \label{dNdp}\\
\mP \frac{\partial N}{\partial \chi_*}&=&
\frac{1}{\sqrt{2\epsilon^\chi_*}}
 \frac{V_*- \Z_\u}{W_*} , \label{dNdc}
\eea
where
\beq
\Z_\u= (V_\u {\epsilon_\u^\phi} - U_\u
{\epsilon_\u^\chi})/\epsilon_\u \,.
 \eeq

To compute the second derivatives we differentiate Eqs.~(\ref{dNdp}) and
(\ref{dNdc}) with respect to $\phi_*$ and $\chi_*$.
This yields
 \bea
 \label{d2Ndp2}
\mP^2 \frac{\partial^2 N}{\partial \phi_*^2} &=&
1
 - \frac{\eta_*^\phi}{2 \epsilon_*^\phi} \frac{U_*+\Z_\u}{W_*} +
 \frac{\mP}{W_*\sqrt{2\epsilon^\phi_*}}
 \frac{\partial \Z_\u}{ \partial \phi_*}
, \\
 \label{d2Ndpdc}
\mP^2 \frac{\partial^2 N}{\partial \chi_*^2} &=& 1 -
\frac{\eta_*^\chi}{2 \epsilon_*^\chi} \frac{V_*-\Z_\u}{W_*} -
 \frac{\mP}{W_*\sqrt{2\epsilon^\chi_*}}
 \frac{\partial \Z_\u}{ \partial \chi_*},
\\
 \label{d2Ndc2}
\mP^2 \frac{\partial^2 N}{\partial \phi_* \partial \chi_*} &=&
 \frac{\mP}{W_*\sqrt{2\epsilon^\phi_*}}
 \frac{\partial \Z_\u}{ \partial \chi_*}
 = -
  \frac{\mP}{W_*\sqrt{2\epsilon^\chi_*}}
 \frac{\partial \Z_\u}{ \partial \phi_*}.
 \eea

In single field inflation, perturbations on large scales are
adiabatic and hence $\zeta$ remains constant during inflation
\cite{Wands:2000dp}. In two-field models, however, the large-scale
perturbations are not in general adiabatic and $\zeta$ evolves
during inflation. In the previous expressions, Eqs.~(\ref{dNdp}),
(\ref{dNdc}) and (\ref{d2Ndp2}--\ref{d2Ndc2}), the function $Z_\u$
takes account of this evolution by expressing how the local
expansion depends upon the local field values on the uniform
density hypersurface during inflation.

\subsection{Curvature perturbation and non-Gaussianity during inflation}
\label{sec:nonG}

We now compute the curvature perturbation and the non-Gaussianity
on a comoving or uniform energy density hypersurface during
slow-roll inflation, as given in Eq.~(\ref{zeta}). From the first
derivatives of $N$, Eqs.~(\ref{dNdp}) and (\ref{dNdc}), one can
derive an expression for the power spectrum of the curvature
perturbation (\ref{Rtophi}). It can be simplified by introducing
two dimensionless variables, related to the values of the
potentials and first slow-roll parameters at $t_*$ and $t_\u$,
\beq
u \equiv \frac{U_*+Z_\u}{W_*}, \qquad v \equiv\frac{V_* -
Z_\u}{W_*}. \label{u_v}
\eeq

The power spectrum then reads
\beq
 \PP_\zeta = \frac{W_*}{24\pi^2 \mP^4}
  \left( \frac{u^2}{\epsilon^\phi_*}
+ \frac{v^2}{\epsilon^\chi_*}
  \right) \,. \label{PP}
 \eeq
Note that this can never be less than the power spectrum
derived by considering only adiabatic perturbations restricted to the
background trajectory (at fixed $C$). In multi-field inflation
there is an additional contribution to the power spectrum
due to non-adiabatic perturbations at Hubble exit.
Thus one has $ \PP_\zeta \geq \PP_{\left. \zeta \right|_C}$,
where $\PP_{\left. \zeta \right|_C}$
can be obtained by taking the limit $t_\u \to t_*$
in Eq.~(\ref{PP}), since for adiabatic perturbations the curvature
perturbation does not change after Hubble crossing.
In this limit
\beq
u \to \frac{\epsilon_*^\phi}{\epsilon_*} , \qquad v
\to \frac{\epsilon_*^\chi}{\epsilon_*} , \qquad {\rm
for } \ \ t_\u \to t_*, \label{limit}
\eeq
and one finds
\beq
\PP_{\left. \zeta \right|_C}
= \frac{W_*}{24\pi^2 \epsilon_*
 \mP^4} \,. \label{PP_zeta_C}
 \eeq

To compute the scalar spectral index $n_\zeta$ with Eq.~(15) 
and the nonlinear
parameter $f_{\rm NL}$ with Eq.~(26) 
one needs the second derivatives $N_{,IJ}$,
given in Eqs.~(\ref{d2Ndp2}--\ref{d2Ndc2}). In the expression
for the scalar spectral index the derivatives of $\Z_\u$ with
respect to the fields contained in these equations mutually
cancel. Thus, by using Eq.~(\ref{si}), one finds
\beq
n_\zeta -1 =-2 \epsilon_* - 4 \frac{u \left(1 -
\frac{\eta_*^\phi}{2 \epsilon_*^\phi}u \right)
     + v \left(1 - \frac{\eta_*^\chi}{2 \epsilon_*^\chi}v \right)
      }{\frac{u^2}{\epsilon_*^\phi}
             +\frac{v^2}{\epsilon_*^\chi}  }
\,.
 \label{si2}
\eeq
Note that in the limit $t_\u \to t_*$, using Eq.~(\ref{limit}), one finds
the spectral index for purely adiabatic perturbations,
\beq
n_{\zeta|_C} -1= -6\epsilon_* +2 \eta^{\sigma \sigma}_*, \label{n_sf}
\eeq
where we have defined \cite{Wands:2002bn}
\beq
\eta^{\sigma \sigma} \equiv
(\epsilon^\phi\eta^\phi+\epsilon^\chi\eta^\chi)/\epsilon.
\eeq
This represents the effective mass of adiabatic fluctuations
{\em tangent} to the background trajectory.

The momentum dependent nonlinear parameter, $f^{(3)}_{\rm NL}$,
defined in Eq.~(\ref{fNL3}), takes a very simple form, derived
from Eq.~(\ref{PP_zeta_C}),
\beq
-\frac{6}{5}  f^{(3)}_{\rm NL}
=
 \epsilon_*\frac{ \PP_{\zeta|_C}}{ \PP_\zeta} (1+f) \leq
 \epsilon_* (1+f),
\label{fNL3bis}
\eeq
where $f$ is the momentum dependence, defined in Eq.~(\ref{f_def}).

In order to give an expression for $f^{(4)}_{\rm NL}$, one needs
$\partial \Z_\u / \partial \phi_*$ and $\partial \Z_\u / \partial
\chi_*$, which can be computed by employing
Eq.~(\ref{dphiIofdphiJ}). One finds
\beq
 \label{dFi}
\sqrt{\epsilon_*^\phi} \frac{\partial \Z_\u}{\partial \phi_*} = -
\sqrt{\epsilon_*^\chi} \frac{\partial \Z_\u}{\partial \chi_*} =
\frac{\sqrt{2}}{\mP} W_* \B,
 \eeq
where we have defined
\beq
\B \equiv - \frac{W_\u^2}{W_*^2} \frac{\epsilon_\u^\phi
\epsilon_\u^\chi}{\epsilon_\u}
 \left(1 - \frac{\eta^{ss}_\u}{\epsilon_\u} \right), \label{B}
\eeq
with \cite{Wands:2002bn}
\beq
\eta^{ss} \equiv
(\epsilon^\chi\eta^\phi+\epsilon^\phi\eta^\chi)/\epsilon.
\label{etass}
\eeq
This represents the effective mass of isocurvature fluctuations
orthogonal to the background trajectory.

We are thus able to give an analytic expression for  $f^{(4)}_{\rm
NL}$ defined in Eq.~(\ref{fNL4}). One finds
\bea
-\frac{6}{5}  f^{(4)}_{\rm NL}&=& 2
\frac{\frac{u^2}{\epsilon_*^\phi}
\left(1
 - \frac{\eta_*^\phi}{2 \epsilon_*^\phi}
u
\right)
+ \frac{v^2}{\epsilon_*^\chi}
\left(1
 - \frac{\eta_*^\chi}{2 \epsilon_*^\chi }
v
\right)
+ \left( \frac{u}{{\epsilon_*^\phi}}
- \frac{v}{{\epsilon_*^\chi}}
\right)^2
\B}{\left( \frac{u^2}{\epsilon^\phi_*}
+ \frac{v^2}{\epsilon^\chi_*} \right)^2}
. \label{fNL2f}
\eea
This is the exact expression for the amplitude of the nonlinear
parameter for the curvature perturbation $\zeta$ during slow-roll
inflation in an arbitrary two-field inflation model with separable
potential, and represents one of the main results of this paper.

As for the power spectrum and spectral index, we can take the
limit of $t_\u \to t_*$, which gives the nonlinear parameter for
purely adiabatic perturbations in the two-field case. Using
Eq.~(\ref{limit}) in Eqs.~(\ref{fNL3bis}) and (\ref{fNL2f}),
yields
 \beq
- \frac{6}{5} f_{{\rm NL}|_C} =
\epsilon_* (1+f) + 2\epsilon_* - \eta_*^{\sigma \sigma}
 \eeq
which coincides with the single field case result
\cite{Maldacena,Allen:2005ye}. Note that, using Eqs.~(\ref{n_sf})
and (\ref{r}) for the adiabatic case, this can be expressed as a
consistency relation between observables \cite{Maldacena},
 \bea
- \frac{6}{5} f_{{\rm NL}|_C} = - \frac{1}{2}(n_{\zeta|_C} -1)+
\frac{r_{|_C}}{8} f \,, \label{consistency}
 \eea
and is thus constrained to be small.

Although the general expression (\ref{fNL2f}) for two fields,
which allows for non-adiabatic perturbations, is rather involved,
we can qualitatively discuss the order of magnitude we expect for
$f^{(4)}_{\rm NL}$. Since ${\rm max}(u,v)\leq 1$ and $u+v=1$, the
denominator in Eq.~(\ref{fNL2f}) is of order $\varepsilon_*^{-2}$,
where we use $\varepsilon$ to denote generic first-order slow-roll
parameters, $\epsilon$ or $\eta$. The first two terms in the
numerator of Eq.~(\ref{fNL2f}) are of order $\varepsilon_*^{-1}$,
so their contribution to $f^{(4)}_{\rm NL}$ is of order
$\varepsilon_*$. Only the third term, which is of order
$\B\varepsilon_*^{-2}$,  leads to a contribution that is not
automatically slow-roll suppressed.

Although $\eta^{ss}$ may become larger than unity during
inflation, the prefactor in front of the parenthesis in
Eq.~(\ref{B}) can be very small or vanishing, suppressing the
contribution of $\B$ to the nonlinear parameter. If this is not
the case, since $\B$ does not appear in the expression for the
spectral index $n_\zeta$, Eq.~(\ref{si2}), for $\eta^{ss}_\u/
\epsilon_\u \gg1$ one may have models with large $f_{\rm NL}$ and
quasi-scale-invariant spectral index, corresponding to large
deviations from the consistency relation for purely adiabatic
perturbations, Eq.~(\ref{consistency}). We leave to future work
the investigation of models where $\B$, and thus $f_{\rm NL}$, are
large.

\section{Perturbations after inflation}

So far we have considered only the curvature perturbation during
inflation. In order to relate our calculations to observables we
need to calculate $\zeta$ {\em after} inflation when the universe
is radiation dominated, on a uniform energy density hypersurface
for $t=t_\u> t_e$, where $t_e$ denotes the end of inflation. In
this case $t_\u$ defines a uniform radiation density hypersurface.

In single field inflation large-scale perturbations are adiabatic
and thus the curvature perturbation on uniform density
hypersurfaces remains constant both {\em during} and {\em after}
inflation, independent of the detailed physics occurring at the
end of inflation. In two-field models we need to take account of
how the local expansion depends upon the local field values both
during inflation and at the end of inflation in order to calculate
$\zeta$ some time after inflation.

If one of the two fields -- for instance $\phi$ -- has stabilized
before the end of inflation, so that the end of inflation is
dominated by a single field $\chi$, then $Z_\u$ becomes constant,
and the power spectrum, the scalar spectral index and the
non-linear parameters after the end of inflation are simply given
by Eqs.~(\ref{PP}), (\ref{si2}), (\ref{fNL3bis}) and (\ref{fNL2f})
with $Z_\u = U_\u=$ const. Note that in this case $\B =0$ at the
end of inflation and the nonlinear parameter $f_{\rm NL}^{(4)}$
will be small.

If both fields are ``active'' at the end of inflation, and
inflation takes place on a hypersurface $\q(\phi_e,\chi_e)=$ const
at $t=t_e$, undefined -- i.e., not necessarily a uniform energy
density, one can separate $\delta N(t_\u, t_*)$ into the sum of
two pieces, the first from $t=t_*$ to the end of inflation
$t=t_e$, the second from $t=t_e$ to a uniform density hypersurface
after inflation $t=t_\u$,
\beq
 \delta N(t_\u, t_* ) = \delta N(t_e, t_*) +\delta N(t_\u, t_e).
 \label{efolds2int}
 \eeq
The first piece, $\delta N(t_e, t_*)$, can be computed by a
similar calculation to that of Sec.~\ref{sec:pesr}. It is possible
to show that $\delta N$ expanded at second order is found by
replacing $Z_\u$ in Eqs.~(\ref{dNdp}), (\ref{dNdc}) and
(\ref{d2Ndp2}--\ref{d2Ndc2}) by
\beq
 \label{Fi}
\Z_e = \left( V_e \frac{\partial \q}{\partial \phi_e}
\sqrt{\epsilon^\phi_e} - U_e \frac{\partial \q}{\partial \chi_e}
\sqrt{\epsilon^\chi_e}\right)  \left( \frac{\partial \q}{\partial
\phi_e} \sqrt{\epsilon^\phi_e} + \frac{\partial \q}{\partial
\chi_e} \sqrt{\epsilon^\chi_e} \right)^{-1}.
 \eeq
Then one needs to compute $\delta N(t_\u, t_e)$.

In scenarios such as the curvaton or modulated reheating
scenarios, it is assumed that the end of inflation hypersurface is
effectively unperturbed, $\delta N(t_e, t_*)\simeq0$, and that the
primordial density perturbation originates from isocurvature field
perturbations, $\delta s$, during inflation, which introduce a
perturbation $\delta N(t_\u, t_e)\propto \delta s$ only {\em
after} or at the {\em end} of inflation.

An alternative possibility is to assume that inflation ends due to
a sudden instability triggered by some function of the fields
reaching a critical value
\cite{Bernardeau:2004zz,Lyth:2005qk,Salem:2005nd}. This is what
happens in hybrid inflation \cite{Linde:1993cn,Copeland:1994vg}
where the false vacuum state is destabilized when the inflaton
field, $\phi$, reaches a critical value, $\phi=\phi_e$. If we
assume instantaneous reheating of the universe one has
\cite{Lyth:2003im}
 \bea
  \delta N(t_\u, t_e , \bx) &=& {\cal N} \left(\rho(t_\u,\bx) \right)
  - {\cal N} \left( \rho(t_e,\bx)
\right) \nonumber \\
&=& - \left[ \frac{d N }{d \rho} \delta \rho + \frac{1}{2}
\frac{d^2 N}{d \rho^2} \delta \rho^2 \right]_e
\nonumber \\
&=& - H_e \left[  \frac{\delta \rho}{\dot \rho} + \frac{1}{2}
\left( \frac{\dot \rho}{2 \rho}  - \frac{\ddot \rho}{ \dot \rho}
\right) \frac{\delta \rho^2}{\dot \rho^2} \right]_e.
 \eea
We assume that the energy density is conserved at the end of
inflation, so that the background value of the energy density in
this expression is $\rho(t_e) = U(\phi_e) + V(\chi_e)$.
Furthermore, if the equation of state becomes radiation-like, so
that $\dot\rho=-4H\rho$, we then have
\begin{equation}
 \delta N(t_\u, t_e , \bx) = \frac{1}{4} \left[ \frac{\delta \rho}{\rho} -
 \frac{1}{2} \frac{\delta \rho^2}{\rho^2} \right]_e \,.
\end{equation}
Note that the numerical coefficient that appears in front of the
brackets in this expression is dependent upon the equation of
state after inflation has ended and would
be $1/3(1+w)$ for an equation of state $P=w\rho$.

In a hybrid-type limit, where the false vacuum dominates the
self-interac\-tion energy of the slowly-rolling fields, we can
take the effective potential to be almost completely flat
(consistent with the slow-roll approximation) so that
$[\delta\rho/\rho]_e$ is negligible and $\delta N(t_\u, t_e)\ll
\delta N(t_e,t_*)$. The primordial curvature perturbation $\zeta$
is then given by the curvature of the end of inflation
hypersurface, $\zeta(t_\u) = \delta N(t_e,t_*)$.

\section{Double quadratic inflation}

To be more specific and give a quantitative estimate of the
non-Gaussianity, it is instructive to
consider the case of massive fields with potential given by
Eq.~(\ref{potential}) with \cite{Polarski:1992dq,Langlois:1999dw}
\beq
U= \frac{1}{2} \mp^2 \phi^2, \quad \quad   V= \frac{1}{2} \mc^2
\chi^2 \,. \label{potential2}
 \eeq
These scalar fields thus have no explicit interaction, but can
interact gravitationally during inflation.

\subsection{Slow-roll analysis}
\label{sec:double}

We can apply our earlier analysis to calculate the curvature
perturbation during slow-roll inflation in this model.

The integrals (\ref{Ndefinite}) and (\ref{integral}) yield
 \bea
 \mP^2 N(t_\u,t_*) &=& \frac14 \left( \phi_*^2 +\chi_*^2 \right)
-\frac14 \left( \phi_\u^2 +\chi_\u^2 \right) \,,  \label{NNN}\\
 \mP^{-2} C(t_\u,t_*) &=& m_\phi^{-2} \ln \left( \frac{\phi_*}{\phi_\u} \right)
 - m_\chi^{-2} \ln \left( \frac{\chi_*}{\chi_\u} \right) \,,
 \eea
where, as for $N$, we have fixed the limits of integration in the
definition of $C$ in Eq.~(\ref{integral}) to run from $t_*$ to
$t_\u$. In the standard treatment, one can parameterize the
slow-roll trajectories of the scalar fields in polar coordinates
\cite{Polarski:1992dq}
\beq
  \chi = 2 \mP \sqrt{N} \sin\theta,
\quad \quad \phi = 2 \mP \sqrt{N} \cos\theta.
 \eeq
The advantage of doing so is that the angular variable
$\theta$ can be related to the number of $e$-foldings by the
expression \cite{Polarski:1992dq,Langlois:1999dw,Gordon:2000hv}
\beq
 N=N_0
\frac{(\sin \theta)^{2/(R^2-1)}}{(\cos \theta)^{2R^2/(R^2-1)}},
\label{Noftheta}
 \eeq
where we have defined the ratio between the masses of the fields
as
\beq
 R={\mc}/{\mp} \,.
  \eeq
Thus, Eq.~(\ref{Noftheta}) implies that an inflationary model is
completely specified by giving $R$ and the values of the fields
$\phi$ and $\chi$ -- or equivalently $N$ and $\theta$ -- at some
given time. Once specified an initial condition,
Eq.~(\ref{Noftheta}) can be used to evolve the background fields
or equivalently the slow-roll parameters. Then one can use
Eqs.~(\ref{PP}), (\ref{si2}), (\ref{fNL3bis})
and (\ref{fNL2f}), to evaluate completely {\em
analytically} the power spectrum $\PP_\zeta$, the scalar spectral
index $n_\zeta$ and the nonlinear parameter $f_{\rm NL}$
for double inflation.

Before studying in detail these equations for a specific case, one
can estimate the amount of non-Gaussianity generally produced {\em
during} and {\em after} inflation. {\em During} inflation, as
explained in Sec.~\ref{sec:nonG}, $f_{\rm NL}$ can be large only
if, in Eq.~(68), the contribution 
proportional to the quantity $\B$ defined in Eq.~(\ref{B}) is
large. In double inflation this contribution can become temporarily large,
at most of order unity, at the turn of the trajectory in field
space.
However, {\em after} inflation, the fields have settled to their
minima, which implies $Z_\u=0$ and also $\B=0$. In this case the
dependence on the field values at $t=t_\u$ in Eq.~(\ref{u_v})
disappears and we can simply rewrite Eqs.~(\ref{PP}), (\ref{si2}),
(\ref{fNL3bis}) and (\ref{fNL2f}) by using the potential
(\ref{potential2}) and Eq.~(\ref{NNN}) with $\phi_\u=\chi_\u=0$,
as
\bea
\PP_\zeta &=&  \frac{H^2_*}{4 \pi^2 \mP^2} N(t_\u,t_*),\\
n_\zeta -1 &=&  -2 \epsilon_* - \frac{1}{N(t_\u,t_*)}, \\
-\frac{6}{5}f_{\rm NL} &=& \frac{1}{2 N(t_\u,t_*)} \left[ 2 +
f(k_1,k_2,k_3)\right] \label{Num},
\eea
where $N(t_\u,t_*)$ is the number of $e$-foldings between Hubble
exit and the moment when $\phi_\u=\chi_\u=0$, which roughly
coincides with the end of inflation. These results coincide with
the estimate of Ref.~\cite{Alabidi}. We conclude that double
inflation {\em cannot} produce large non-Gaussianity -- i.e.,
$|f_{\rm NL}| \ll 1$ -- these being suppressed by the slow-roll
conditions on the fields {\em at Hubble exit}. We now turn to a
more detailed analysis of Eqs.~(\ref{PP}), (\ref{si2}),
(\ref{fNL3bis}) and (\ref{fNL2f}), and compare with a numerical
results for the evolution on large scales without assuming
slow-roll.

\subsection{Numerical analysis}
\label{sec:numerical}

The $\delta N$-formalism  assumes that the fields are in slow-roll
at Hubble exit, Eq.~(\ref{srdN}) (see \cite{Lee:2005bb} for a
development of the $\delta N$-formalism applicable to more general
situations). Furthermore, in deriving Eqs.~(\ref{PP}),
(\ref{si2}), (\ref{fNL3bis}), (\ref{fNL2f}) and the background
evolution relation (\ref{Noftheta}), we have assumed slow-roll all
along the inflationary evolution, from $t_*$ to $t_\u$.

While still assuming that slow-roll conditions are satisfied at
Hubble exit, one can go {\em beyond} the slow-roll approximation
by {\em numerically} solving the full background evolution
equations for the fields, Eqs.~(\ref{evol_phi}) and
(\ref{evol_chi}), together with the Friedmann equation
(\ref{fried_1}), and computing the evolution of the scale factor
$a$. One can then evaluate the expansion $N=\ln a$ as a function
of the field initial values $(\phi_*,\chi_*)$, up to some final
time $t_\u$ which may be some time {\em during} or {\em after}
slow-roll inflation. Then, one can numerically calculate the first
and second partial derivatives of $N(t_\u,t_*)$ with respect to
$\phi_*$ and $\chi_*$ by the {\em finite-difference method}
\cite{nr}, i.e., by computing $N$ for different values close to
$(\phi_*,\chi_*)$. Indeed, this provides an efficient method to
compute the non-Gaussianity numerically for {\em any} two-field
model, also when the field potential is not separable.

\begin{figure}
\begin{center}
\includegraphics[width=3.5in]{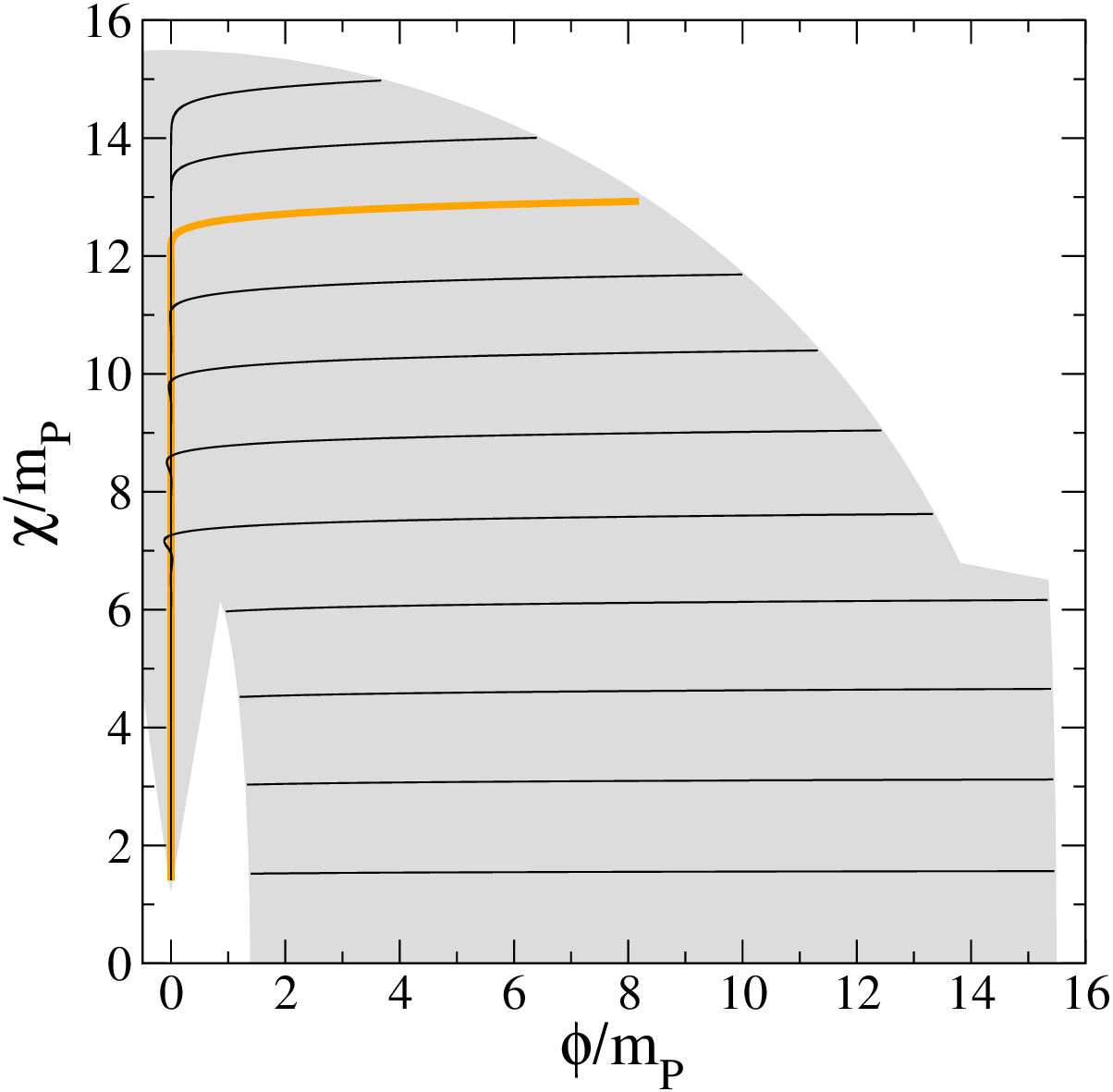}
\end{center}
\caption{Examples of trajectories in field space are shown from
Hubble exit, starting $60$ $e$-foldings before the end of
inflation at $\epsilon_e=1$, for $R=1/9$. The thick (orange) line
represents the trajectory starting from $\phi=\chi=13 \mP$, and
shown from Hubble exit, $(\phi_*=8.2,\chi_*=12.9)$, to the end of
inflation, $(\phi_e=0.0,\chi_e=1.4)$. The grey shading represents
the space of all possible trajectories.} \label{fig:1}
\end{figure}

\begin{figure}
\begin{center}
\includegraphics[width=3.8in]{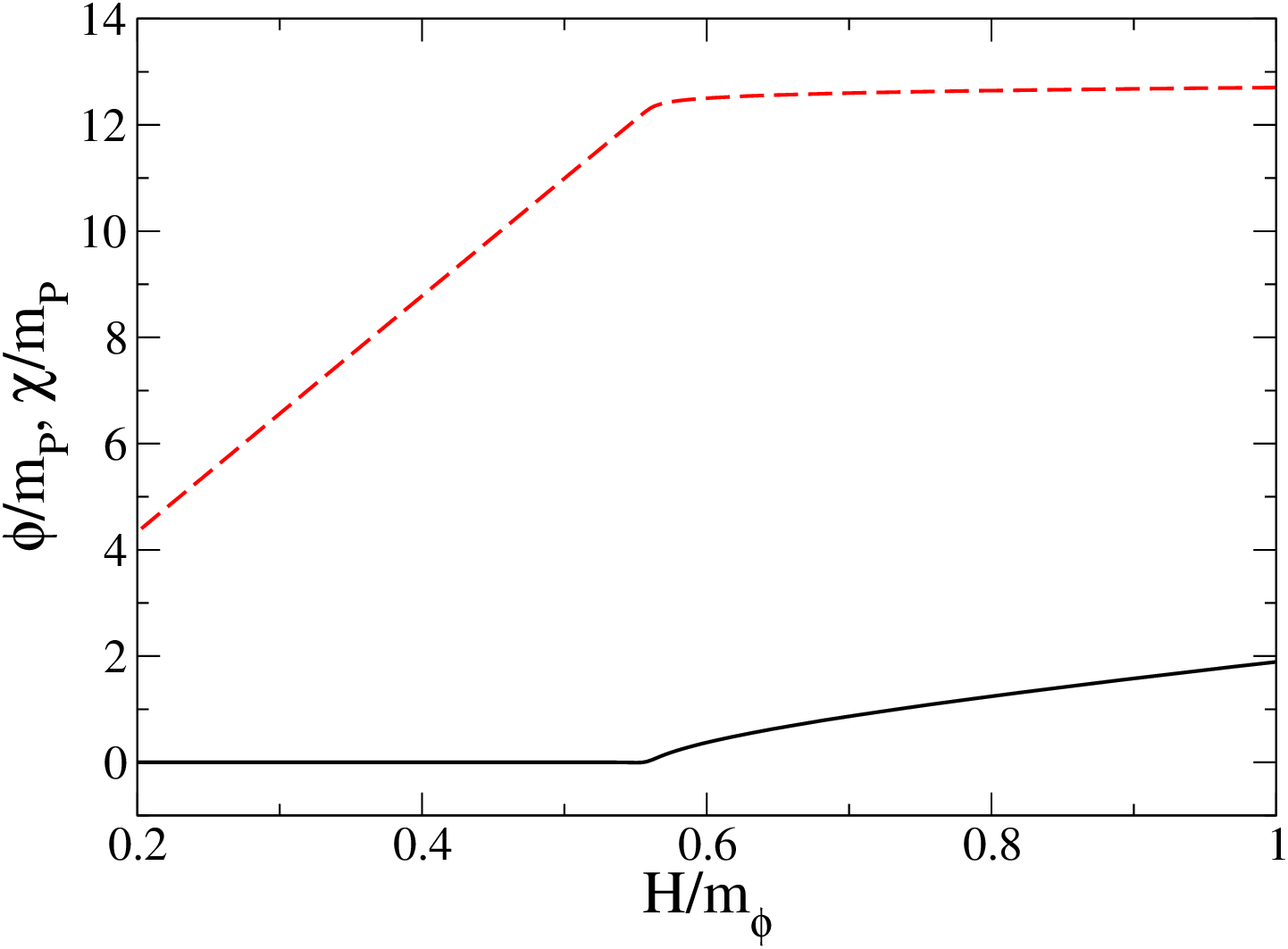}
\end{center}
\caption{The values of the fields $\phi$ (solid line) and $\chi$
(dashed red line)
during the inflationary trajectory of Fig.~\ref{fig:1}, are shown
as a function of the Hubble rate $H$, from $N\simeq 42$ to $N
\simeq 4$ $e$-foldings from the end of inflation. Note that time increases
from right to left. } \label{fig:2}
\end{figure}

\begin{figure}
\begin{center}
\includegraphics[width=3.8in]{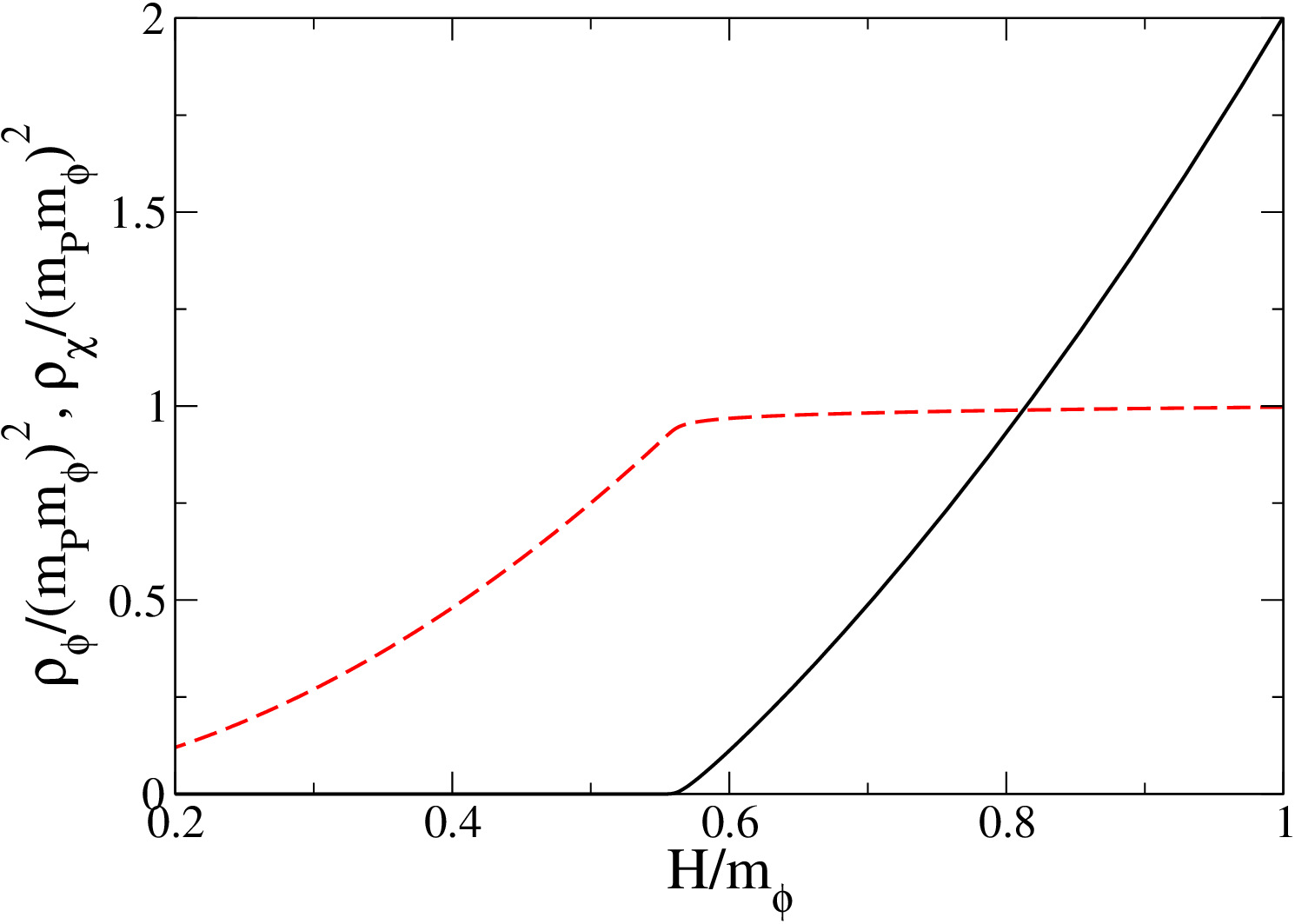}
\end{center}
\caption{ The energy densities of the fields $\rho_\phi $ (solid
line) and $\rho_\chi $ (dashed red line) normalized to $(\mP
m_\phi)^2$, are shown as for Fig.~\ref{fig:2}.} \label{fig:3}
\end{figure}

\begin{figure}
\begin{center}
\includegraphics[width=3.8in]{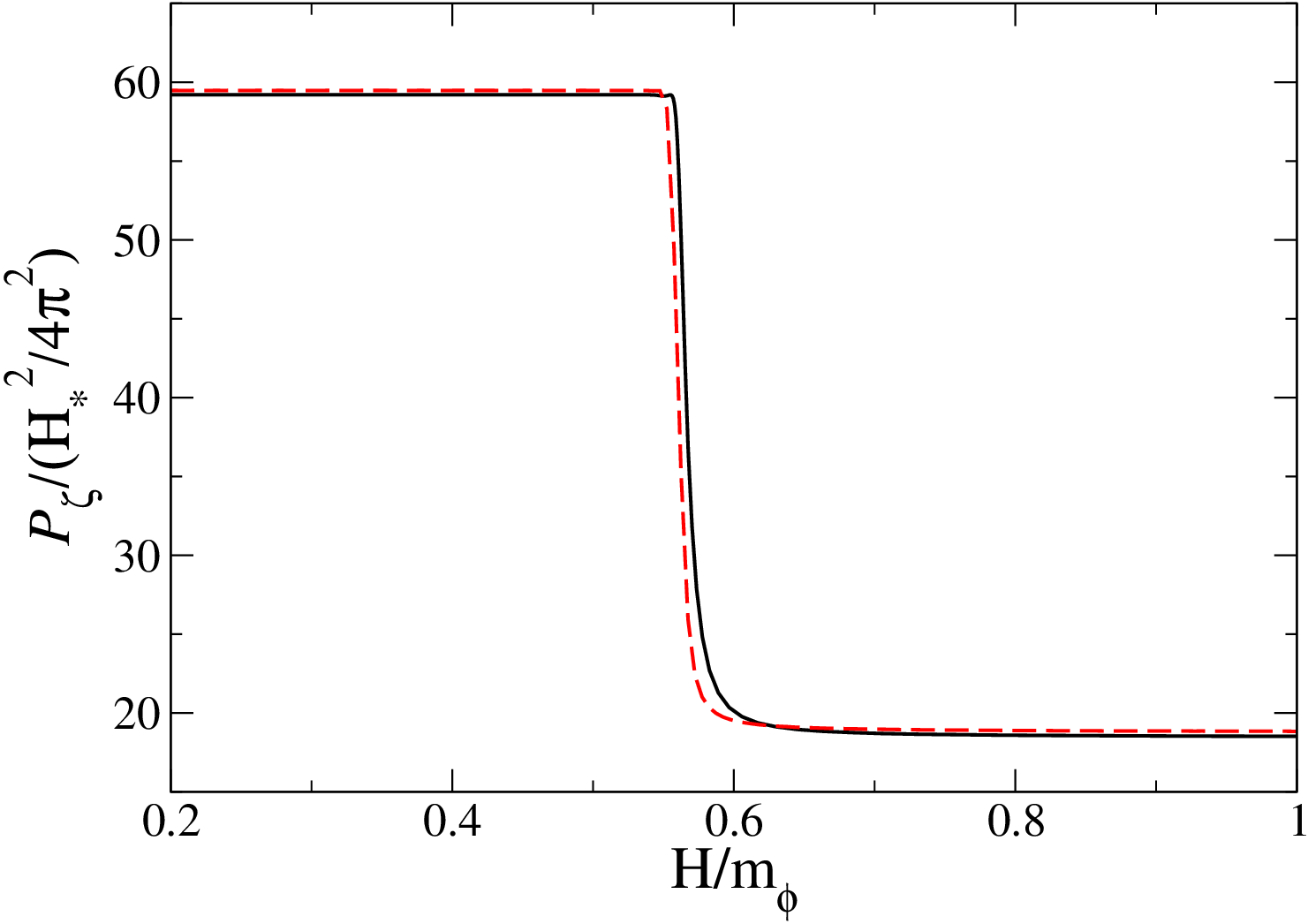}
\end{center}
\caption{The power spectrum $\PP_\zeta$ of the large-scale uniform
density perturbation $\zeta$ during the inflationary trajectory of
Fig.~\ref{fig:1}, is shown as a function of the Hubble rate $H$,
from $N\simeq 42$ to $N \simeq 4$ $e$-foldings from the end of
inflation ($\mP=1$). The solid and the dashed lines represent the analytic
and numerical calculations, respectively.} \label{fig:4}
\end{figure}

\begin{figure}
\begin{center}
\includegraphics[width=3.8in]{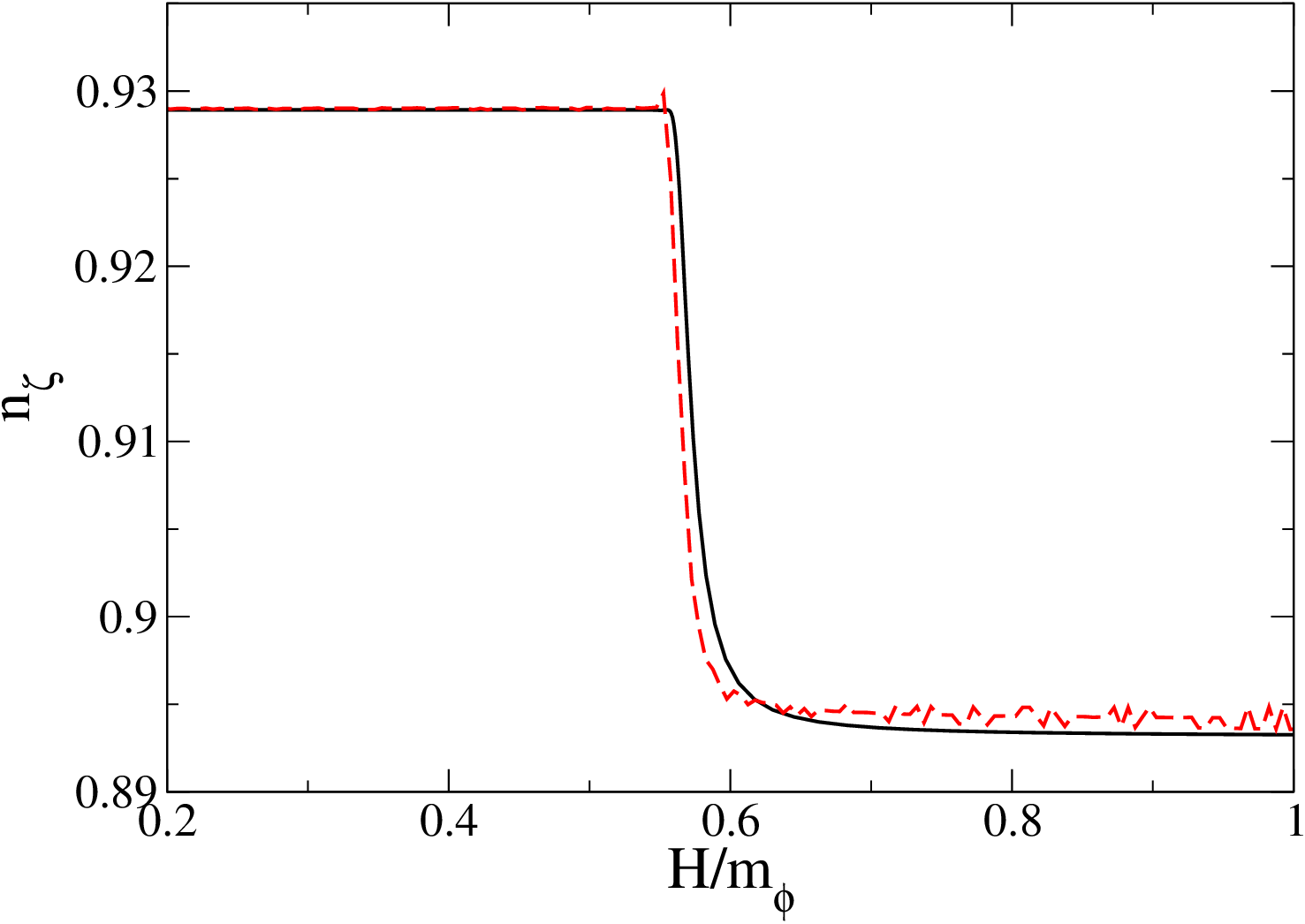}
\end{center}
\caption{The spectral index $n_\zeta$ is shown as for
Fig.~\ref{fig:4}.} \label{fig:5}
\end{figure}

\begin{figure}
\begin{center}
\includegraphics[width=3.8in]{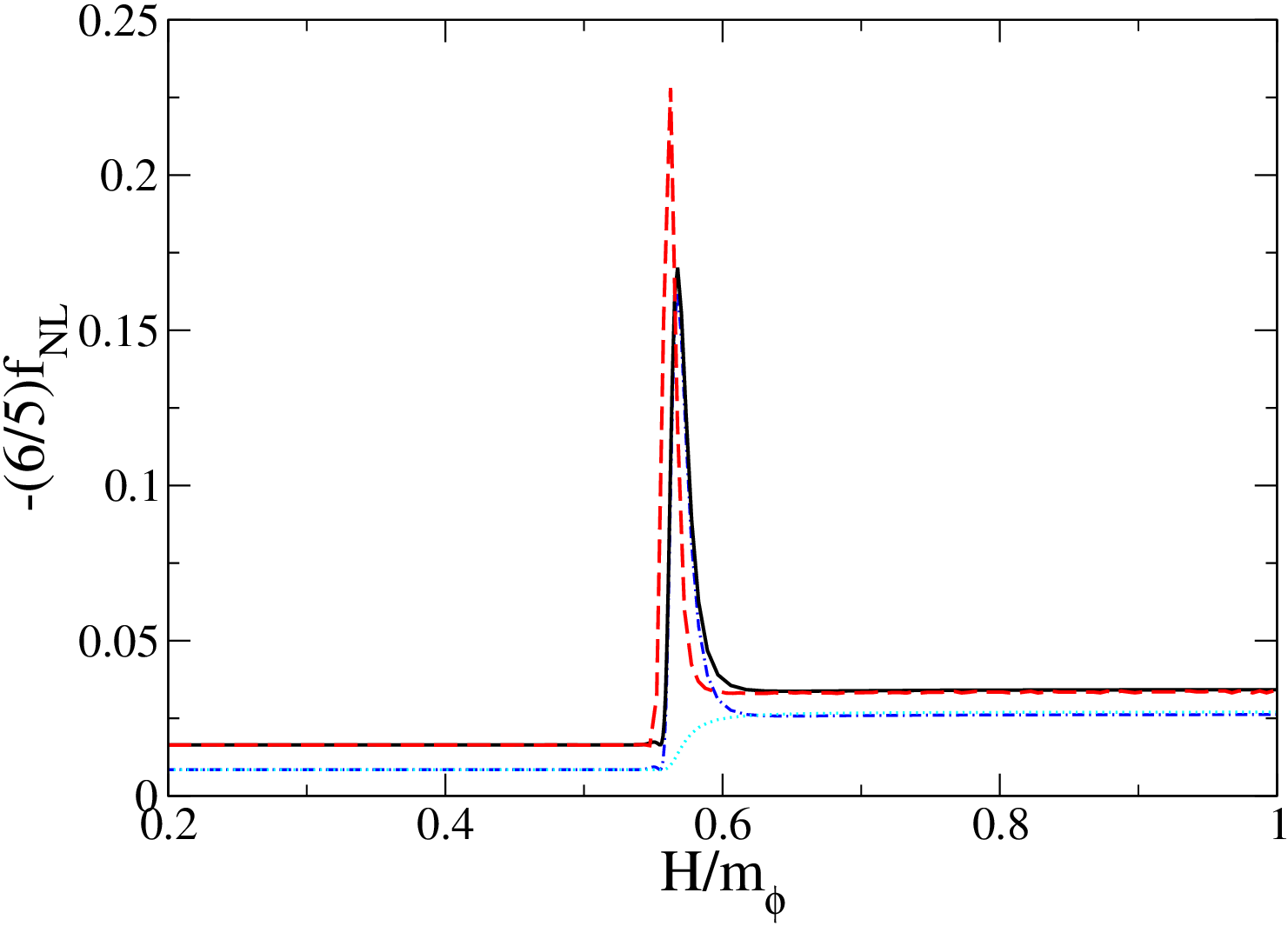}
\end{center}
\caption{The nonlinear parameter $f_{\rm NL}$, in the limit where
$f(k_1,k_2,k_3)=0$ ($k_1 \ll k_2 \approx k_3$) is shown as for
Fig.~\ref{fig:4}. The (turquoise) dotted line represents
$-(6/5)f_{\rm NL}^{(3)}/(1+f)$ while the (blue) dashed-dotted line
represents $-(6/5)f_{\rm NL}^{(4)}$, both computed analytically.}
\label{fig:6}
\end{figure}

We study, as an example, the specific case considered by
Rigopoulos {\em et al.}~in \cite{Rigopoulos:2005us}: mass ratio
$R=1/9$ and initial condition $\phi=\chi=13 \mP$. We assume that
inflation ends on a hypersurface $\epsilon_e = 1$ and that Hubble
exit takes place at $t=t_*$, corresponding to $60$ $e$-foldings
before the end of inflation. One finds that $(\phi_e= 0.0,
\chi_e=1.4)$, corresponding to $\theta_e=\pi/2$, and that $(\phi_*
= 8.2,\chi_* = 12.9)$, corresponding to $\theta_*=0.32 \pi$.

The trajectory in field space is shown in Fig.~\ref{fig:1}. One
can see that assuming slow-roll from $t_*$ to $t_e$ is well
justified. Indeed, the {\em turn} of trajectory in field space,
corresponding to the moment when $\phi$ exits slow-roll, takes
place near $\phi=0$, when the energy density of $\phi$ has already
become negligible. This can also be checked in Figs.~\ref{fig:2}
and \ref{fig:3}, where
the values of the fields and their energy densities are shown,
respectively, at the turn of the trajectory in field space, from
$N\simeq 42$ to $N \simeq 4$ $e$-foldings from the end of
inflation.

Now we compute $\PP_\zeta$, $n_\zeta$ and $f_{\rm NL}=f_{\rm
NL}^{(3)}+f_{\rm NL}^{(4)}$ analytically, using Eqs.~(\ref{PP}),
(\ref{si2}), (\ref{fNL3bis}) and (\ref{fNL2f}), and numerically,
using the procedure explained above. In
Figs.~\ref{fig:4}--\ref{fig:6} we compare the analytic and
numerical calculation of the power spectrum $\PP_\zeta$, the
scalar spectral index $n_\zeta$, the nonlinear parameter $f_{\rm
NL}$, as a function of the Hubble rate during inflation. The two
calculations agree remarkably well, within an accuracy that
depends on the numerical precision.

As expected, when the heavy field $\phi$ dominates the total
energy density, for $H/m_\phi \gsim 0.8 $, the uniform density
curvature perturbation $\zeta$ remains constant. It starts growing
when $H$ drops to
$0.55 \lsim H/m_\phi \lsim 0.8$, corresponding to  both the light and
heavy fields contributing to the total energy density. During this
phase, the large scale total entropy perturbation is
non-vanishing, and sources the evolution of $\zeta$
\cite{Garcia-Bellido:1995kc}. After few oscillations, the heavy
field energy density is red-shifted and the light field dominates
the universe while $\zeta$ becomes constant in time again.
Consequently, the power spectrum and the spectral index grow
monotonically, following the growth of $\zeta$.

The evolution of the nonlinear parameter $-f^{(4)}_{\rm NL}$,
however, is non-mo\-no\-to\-nic. It grows sharply during the
intermediate phase $0.55 \lsim H/m_\phi \lsim 0.8$, corresponding
to the heavy field leaving slow-roll, but it then decreases. This
growth corresponds to the growth of $\B$ defined in Eq.~(\ref{B}).

One can compute the value of $f_{\rm NL}$ at the end of inflation,
for $\epsilon_e \simeq 1$, in the limit where $f(k_1,k_2,k_3)=0$
($k_1 \ll k_2 \approx k_3$). One finds $-(6/5)f_{\rm NL}^{(4)} =
0.008$ and $-(6/5)f_{\rm NL}^{(3)} = 0.008$ which coincide with
the analytical result of Eq.~(\ref{Num}), $-(6/5)f_{\rm NL} =0.016
\simeq 1/60$. We conclude that $f_{\rm NL}$ in this model is much
smaller that unity.

Note that, as shown in Fig.~\ref{fig:4}, the curvature
perturbation becomes constant after the energy density of the more
massive field ($\phi$ in our case) becomes negligible. The
isocurvature perturbations in the massive field are suppressed at
the end of inflation and hence the nonlinear parameter that we
have calculated is indeed the primordial $f_{\rm NL}$ constrained
by observations, unless we have some curvaton-type mechanism which
alters the large-scale curvature perturbation after inflation.

\section{Conclusion}

In this work we have studied the non-Gaussianity, in terms of the
bispectrum, generated by models of inflation with two fields evolving
during inflation. We have first reviewed the use of the $\delta
N$-formalism to compute the curvature perturbation on uniform density
hypersurfaces during inflation, up to second order in the
perturbations, when the slow-roll conditions are satisfied by all
fields.

We have then specialized to the two-field case, and assumed that
the potential is separable into the sum of two potentials, each of
which is dependent on only one of the two fields. In this case the
number of $e$-foldings can be analytically expanded up to second
order in the initial field fluctuations and we have derived an
analytic formula, Eq.~(\ref{fNL2f}), to compute the nonlinear
parameter $f_{\rm NL}$ in terms of the values of the potential and
slow-roll parameters of the fields. Using this formula, one can
identify the conditions for the model to generate large
non-Gaussianity. In particular, we have shown that a large $f_{\rm
NL}$ after inflation requires a large $\eta^{ss}$ -- defined in
Eq.~(\ref{etass}) -- at the end of inflation, although a large
$\eta^{ss}$ does not necessarily imply large non-Gaussianity.

As a specific model, we have considered  double inflation with two
massive scalar fields. In this case, the background evolution of
the fields can be computed analytically and we have been able to
compute the evolution of the nonlinear parameter during inflation.
As shown in Sec.~\ref{sec:double}, one expects the non-Gaussianity
generated in double inflation to be slow-roll suppressed, $|f_{\rm
NL}| \ll 1$, as in the single field case. Since our analytic
formula relies on the slow-roll conditions of all fields from
Hubble exit until the end of inflation, we have extended our
analysis by developing a numerical method, based on the $\delta
N$-formalism, to compute the curvature perturbation up to second
order in the perturbations, that allows us to relax the slow-roll
assumption after Hubble exit. This method, which can be extended
to any two-field model, confirms our analytic results.

Our results agree with a previous discussion of non-Gaussianity in
a double quadratic potential by Alabidi and Lyth \cite{Alabidi},
based on an estimate of $\delta N$ given in \cite{Lyth:1998xn}
using the integral (\ref{Ndefinite}). In Ref.~\cite{Lyth:1998xn}
it is argued that $N$ in Eq.~(\ref{Ndefinite}) is dominated by the
field values at Hubble exit and hence one can neglect the
dependence of $N$ upon the field values at the final time, $t_\u$.
This is not in general true when evaluating the perturbation
$\zeta$ during inflation. Including the dependence of $\phi_\u$
and $\chi_\u$ upon the field values at Hubble exit considerably
complicates our analysis, but allows to follow perturbations
during inflation. In our numerical example we see that the
non-linearity parameter does evolve during inflation, especially
as the trajectory turns the corner in field space, reflecting its
dependence upon the final field values. However this does not
alter the general conclusion that the non-linearity parameter
remains small at the end of inflation.

Our results disagree with those of Rigopoulos {\em et
al.}~\cite{Rigopoulos:2005us} who found that double quadratic
inflation, in particular the model studied in
Sec.~\ref{sec:numerical}, can generate large non-Gaussianity. The
formalism used by Rigopoulos {\em et al.}~differs from the $\delta
N$-formalism in two aspects: First, it takes into account the
sub-Hubble evolution of the field perturbations via stochastic
terms. Second, it integrates {\em explicitly} the coupled
evolution of the adiabatic and entropy super-Hubble perturbations.
However, we have explicitly calculated in this model the
non-Gaussianity due to the three-point function for the field
perturbations at Hubble exit, using the result of Seery and Lidsey
\cite{Seery}, and shown that it is small, as previously argued by
Lyth and Zaballa \cite{Zaballa}. Furthermore, the $\delta
N$-formalism, on super-Hubble scales, is equivalent to integrating
the evolution of $\zeta$.  Thus, the contribution to $f_{\rm NL}$
coming from the super-Hubble evolution should be equally taken
into account by both formalisms and we are unable to explain the
discrepancy between the results.

We conclude, on the basis of our analysis, that nonlinear
evolution on super-Hubble scales during double quadratic inflation
does not appear to be capable of generating a detectable level of
non-Gaussianity in the bispectrum. However, the possibility of
producing a large non-Gaussianity is left open in models of
two-field inflation where the mass of the isocurvature field
orthogonal to the classical trajectory becomes larger than the
Hubble rate and both the first slow-roll parameters of the two
fields do not vanish at the end of inflation.
\\

{\em Note added:} After completing this work we have learnt
\cite{EPSS} that our conclusion that $f_{\rm NL}$ is small for the
specific double quadratic inflation model investigated in Sec.~5
is in qualitative agreement with an improved numerical calculation
by the
authors of \cite{Rigopoulos:2005us}.\\

{\bf Acknowledgments:} The authors are grateful to Kari Enqvist,
Soo Kim, Andrew Liddle, David Lyth, Gerasimos Rigopoulos, Paul
Shellard, Antti V\"aihk\"onen and Bartjan van Tent for interesting
discussions and David Lyth and Paul Shellard for comments on a
draft of this paper.

\end{document}